\documentclass{PoS}

\title{Confining vacua and Q-state Potts models with Q<1}

\ShortTitle{Confining vacua and Q-state Potts models with Q<1}

\author{\speaker{Ferdinando Gliozzi}\\
        Dipartimento di Fisica Teorica, Universit\'a di Torino and\\
        INFN, Sezione di Torino, Italy\\
        E-mail: \email{gliozzi@to.infn.it}}

\abstract{
In most Yang-Mills models the vacuum where magnetic monopoles condense  
coincides with that where center vortices percolate, thus it is not 
clear which of these two properties is most directly involved in producing 
confinement. It is pointed out that  there is a class of 3D gauge models, 
which can be though of as duals of  Q-state Potts models with Q < 1,  where
the magnetic monopole condensation is a necessary but not sufficient condition
 for percolation of center vortices. A set of numerical tests at 
$Q=\frac1{10}$ shows that there is a vacuum in which the magnetic monopole 
condensate does not yield confinement, in the 
sense that large Wilson loops obey a perimeter law. In such a vacuum the 
center vortices form a dilute gas of loops. 
At stronger coupling  there is also a truly  confining vacuum  where both  
confining mechanisms are present. 
}
\FullConference{The XXV International Symposium on Lattice Field Theory\\
                 July 30 - August 4 2007\\
                 Regensburg, Germany}

\usepackage{amsmath}
\usepackage{amssymb} 
  
\newcommand{\bra}{\langle}
\newcommand{\eq}{\begin{equation}}
\newcommand{\en}{\end{equation}}
\newcommand{\ket}{\rangle}

\def\one{{\rm 1\kern -.9mm l}}

\begin{document}

\section{Introduction}
 Magnetic monopoles \cite{mm} and center vortices \cite{cv} are widely believed to be the 
most important degrees of freedom for confinement in 
Yang Mills theories. 

Plausibility arguments suggest that magnetic 
monopole condensation implies a dual Meissner effect which pushes out of the 
vacuum the colour field and gives the well-known physical picture of confinement in terms of dual Abrikosov vortices describing the confining strings joining the quark sources. Considerable evidence for this dual Meissner effect has been accumulated on the lattice \cite{st}, including the definition of a 
disorder parameter demonstrating the condensation of magnetic monopole 
below the deconfining temperature \cite{dgp}

Center vortices are string-like excitations formed out of the center of the gauge group    which are expected to encode all the infrared physics of confinement
 \cite{Langfeld:1997jx,Del Debbio:1998uu}.  When they 
percolate \cite{Engelhardt:1999fd}, produce  a very efficient disordering
   mechanism which could lead to the area law decay of large Wilson loops.

In most YM models the phase with magnetic monopole condensation coincides
with that where center vortices percolate, thus it is not clear which of these
two properties is most directly involved in producing confinement, or, to be more precise, the area law decay of large Wilson loops.
There are many open, intertwined questions about the validity of these confinement mechanisms. In particular, is it possible to derive monopole condensation from percolation of center vortices or vice versa? Are both mechanisms necessary for confinement? are they also sufficient? 

In this talk I try to answer these questions by studying a particularly simple class of 3D gauge models, which can be though of as duals of  Q-state Potts 
models. In these models the confining mechanisms can be easily identified 
in some specific geometric properties of the random graphs associated 
to the configurations of the Q-state Potts models. In particular it will be evident that the percolation of center vortices implies the condensation of magnetic monopoles. On the contrary, it is pointed out that when 
 Q < 1
the magnetic monopole condensation is not necessarily associated to the 
percolation of center vortices: it is demonstrated through a numerical 
experiment that there is a vacuum state in which, although 
the magnetic monopoles condense,  quark sources are not confined, because  
large Wilson loops  decay exponentially with the perimeter  
instead of the area.
It is also shown that in this theory there is a confining vacuum only 
when the magnetic monopole condensation is associated to the percolation 
of center vortices.  

The contents of this contribution are as follows.
 In the next Section the main properties of the Q-state Potts model and of 
its gauge dual are described with a particular emphasis on the definition 
of Wilson loop, which plays an essential role in the studies of confinement. In the following Section the nature of the vacua of these gauge duals in the range
$0<Q<1$ and the salient features of the phase diagram are discussed. In 
Section 4 a local Monte Carlo algorithm for simulating Potts models in the 
range $0\le Q< 1$ is presented. Finally, in the last Section, some numerical 
results generated with this algorithm are reported.
\section{Q state Potts model and its dual}
The Hamiltonian of the (ferromagnetic) Q-state Potts model is, for Q  integer, 
$H=-\sum_{\bra i\,j\ket}\delta_{\sigma_i\,\sigma_j}$, where the site variable $\sigma_i$ takes the values $\sigma_i=1,2,\dots,Q$, with $\bra i\,j\ket$ ranging over the links of an arbitrary lattice or graph $\Lambda$. This model is symmetric under $S_Q$, the group of permutations of $Q$ elements. The canonical 
partition function $Z=\sum_{\{\sigma\}}e^{-\beta \,H}$ can be rewritten
in the  Fortuin Kasteleyn (FK) random cluster representation:
\eq
Z=\sum_{G\subseteq\Lambda}w_G=\sum_{b,c}\Omega(b,c)\,v^{b}Q^{c}~~~,
\en
where $v=e^\beta-1$ and the summation is over all spanning subgraphs 
of $\Lambda$, $w_G=v^b\,Q^c$ is their weight expressed in terms of the 
number $b$ of edges of $G$, called bonds, and the number $c$ of connected 
components, called FK clusters;  $\Omega(b,c) $ is their multiplicity.
 This representation now defines a model for any real or complex  
$Q$, which acts as the fugacity controlling the number of FK clusters.

When $\Lambda $ is a three-dimensional lattice,
the FK random cluster representation is also useful to define a gauge dual 
of this spin model (for any complex $Q$) . The gauge dual lives in the dual 
lattice $\tilde\Lambda$. 
The most basic observables of any gauge model are the Wilson 
loops. In the present case these are associated to the closed paths  
$\gamma\in\tilde\Lambda$. For any spanning subgraph $G\subseteq\Lambda$ 
the Wilson loop $W_\gamma$ measures the topological entanglement between $\gamma$ and $G$. More precisely we attribute to the Wilson loop
$W_\gamma(G)$ the value 1 if no FK cluster of $G$ is linked to $\gamma$, 
otherwise we set $W_\gamma(G)=0$. The vacuum expectation value of $W_\gamma$ is defined accordingly:
\eq
\bra W_\gamma\ket=\sum_{b,c}W_\gamma(G)\Omega(b,c)\,v^{b}Q^{c}/Z~.
\en 
In the special cases where $Q=2,3,\dots$ this definition coincides with the one obtained by applying the usual Kramers-Wannier duality, provided one defines the topological linking as a winding modulo $Q$   \cite{Gliozzi:1996fy}. The gauge theory  dual to  
$Q=1$ Potts model, corresponding to random percolation, has been  studied in detail in \cite{Gliozzi:2005ny}. In particular it has been shown that, although the gauge group is trivial, it behaves like a full-fledged gauge theory with a 
confining vacuum (corresponding to the percolating phase), a string tension having a well-behaved continuum limit, a non trivial glue-ball spectrum
\cite{Lottini:2005ya} and a deconfinement transition at a well determined temperature. In this talk I describe some new features of this kind of gauge models 
in the range $0<Q<1$.
\section{Confining vacua}
Wilson loops provide us with a fundamental tool for a precise definition of 
confinement in a pure gauge theory. A confining phase is expected 
to show up in an area law decay for the vacuum expectation value of large 
Wilson loops $\bra W_\gamma\ket$. This exactly means that if $\gamma$ 
is scaled up keeping its shape fixed and increasing the
area $A$ of the encircled minimal surface, then 
$\bra W_\gamma\ket\propto\,e^{-\sigma\,A}$, where $\sigma$
defines the string tension. 

For a generic  $Q>0$ one expects two kinds of vacua, depending on the 
value of $\beta$. When $\beta $ is small enough, the system is in a 
symmetric vacuum,
characterised by the formation of FK clusters of finite size. If one 
probes this vacuum with Wilson loops of size much larger than the typical 
dimension of these loops, one finds a perimeter law decay 
$\bra W_\gamma\ket\propto\,e^{-p\,\vert\gamma\vert}$ ($\vert\gamma\vert$ is
the perimeter of the loop), the reason being that the only clusters that can be felt by $W_\gamma$ are those near the closed path $\gamma\in\widetilde\Lambda$.
When $\beta$ is larger than a threshold value $\beta_t$ which depends on the 
kind of lattice $\Lambda$, the $S_Q$ symmetry of the model is 
spontaneously broken 
and the corresponding vacuum is characterised by the formation of an infinite, 
percolating, 
FK cluster $G_\infty\subset G$. The spin field $\sigma_i$ associated to the 
sites of the lattice $\Lambda$ is , in all respects, the disorder parameter of 
the dual gauge theory and the formation of an infinite, percolating, cluster 
is a direct sign of the condensation of magnetic monopoles, 
i.e.$\bra\sigma_i\ket\not=0$. 
It is clear that in this case the number of 
paths of $G_\infty$ piercing the minimal surface encircled by $\gamma$ grows 
with its area $A$, therefore one is tempted to argue that large Wilson loops
obey an area law. Note however that the probe $W_\gamma$ feels only 
those piercing paths which are \underline{closed}, thus in order to conclude 
for a confining vacuum one has to assume that also the subgraph $\cal{C}$ 
composed by the circuits of $G$ has got an infinite component 
$\cal{C}_\infty\subset \cal{C}$ for $\beta\ge \beta_t$. This has been 
demonstrated through numerical simulations only when $Q\ge1$
\cite{Gliozzi:2002ht,Gliozzi:2005ny}. For lesser 
values this is not necessarily true. 

Actually there is a very simple argument showing that, 
keeping constant the mean
number of bonds $\bra b \ket$, the  size   of 
$\cal{C}$ or, more precisely, the number $b_{\cal{C}}$ of bonds belonging to 
$\cal{C}$ is  a 
decreasing function of $Q$ and vanishes in the limit $Q\to0$. In fact $Q$ is 
the cluster fugacity of the system: when $Q$ decreases, so does the number 
of clusters $c$. 
The only way to reduce $c$ is to add {\sl bridges}, i.e. bonds that join
otherwise disconnected clusters. Now the total number of bonds 
$b$ is the sum of $b^{~}_{\cal{C}}$ bonds 
belonging to $\cal{C}$ (i.e. to circuits) and of $b^{~}_{\cal{B}}$  
bridges, i.e. $ b=b^{~}_{\cal{C}}+
b^{~}_{\cal{B}}$,    where $G=\cal{C}\cup\cal{B}$; 
therefore a growth of the bridges
keeping $b$ constant implies decreasing of $b^{~}_{\cal{C}}$, q.e.d.

We note, as a side remark, that there is a general relationship between the 
bonds of the two kinds:
\eq
\bra b^{~}_{\cal{B}}\ket\frac{v+Q}v+\bra b^{~}_{\cal{C}}\ket\frac{v+1}v=N~;
\en
this is true for any Q-state Potts model  on an arbitrary graph with $N$ links
\cite{Caselle:2000yy}.
In the special case of two-dimensional, infinite square lattice at the 
transition point $(v=\sqrt{Q})$ the self-duality of the model requires 
$\bra b\ket=\frac N2$, thus at criticality we have 
$b^{~}_{\cal{C}}=N\frac{\sqrt{Q}}{2(\sqrt{Q}+1)}$. This exact result 
can be used as a check of the Monte Carlo algorithm described in the next 
Section.
\begin{figure}
\centering{\includegraphics[width=8.cm]{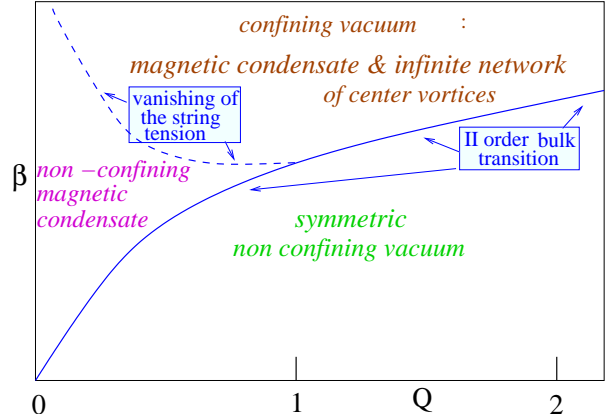}}
\caption{A schematic view of the phase diagram of gauge Q-state Potts model.
The solid line denotes the bulk transition corresponding to the condensation of magnetic monopoles. The dashed line, corresponding to the vanishing of the string tension, does not imply any bulk transitions. }
\end{figure}

The reduction of closed paths as $Q$ decreases suggests, for $Q$ 
\underline{and} 
$\beta-\beta_t$ small 
enough, that the subset $\cal{C}$ does not longer percolate even in the
phase where the symmetry is spontaneously broken. For instance, in the limit
$Q,v\to0$ with the ratio $w=v/Q$ held fixed \cite{qzero} the surviving configurations are spanning subgraphs not containing any circuits, i.e. 
$\cal{C}=\emptyset$, hence in a 3D lattice $W_\gamma(G)=1\,,\,\forall~ 
\gamma$ and $G$ and the $Q=0$ dual gauge theory is trivial.

The above remarks suggest that when $Q<1$ is small enough, the 
standard non-confining vacuum, 
corresponding to the symmetric phase of the Potts model, is 
separated from a truly confining vacuum, where $\cal{C}$, the subgraph 
of circuits,  holds an infinite component, by an 
intermediate vacuum characterised by the condensation of 
magnetic monopoles (hence by the formation of an infinite, 
percolating FK cluster) which however is not confining, because 
the closed paths, corresponding to center vortices, form a dilute 
gas of loops embedded in the infinite cluster 
rather than a connected skein. A sketch of the expected 
phase diagram of the 3D Potts model in the small $Q$ region  is drawn in 
Fig.1.

\section{A  Monte Carlo algorithm for Q<1 Potts models }
The non-local cluster algorithm of Swentsen and Wang \cite{sw} and its 
generalisation to non integer $Q$ \cite{cm} is applicable only for $Q\ge 1$.
In order to study the region $0<Q<1$
 we are interested in we have to resort to some local Monte Carlo algorithm
\cite{s,Gliozzi:2002ub}. I describe here a variant of the method described in
 \cite{ Gliozzi:2002ub} 
which can be  implemented in an efficient way and works only 
when $0\leq Q\leq1$. 

First, divide the interval $[0,1]$ in three parts
$a\,,b-a\,,1-b$, where $a$ and $b\geq a\ge 0$ are suitable functions 
of $\beta$ and $Q$, to be determined later. 
Then apply the following recursive procedure that  generates a Markov 
sequence of  spanning subgraphs $\dots\to G^{(n)}\to
G^{(n+1)}\to \dots  $ of an arbitrary lattice $\Lambda$: 

{\sl i)}  Pick a link $\ell\in\Lambda$ and draw a uniformly 
distributed random number $0\le r_\ell\le 1$; 

{\sl ii)} create or erase  a bond on the 
link $\ell$ according to the following rules: {(a)} put a bond if 
$r_\ell\le a$; (b) erase any bond if  $r_\ell\ge b$; (c) in the 
remaining case $(a<r_\ell<b)$ put a bond only if it is a bridge, i. e. 
only if it connects two otherwise disjoint clusters. 

The Markov chain generated in this way forms an ergodic trajectory in the space of  configurations of the Q-state Potts model. Requiring detailed balance 
with respect the equilibrium distribution yields 
$a=1-e^{-\beta}\equiv p$ and $ b=\frac p{Q\,(1-p)+p}$. The inequality $a\le b$
implies $Q\le1$\footnote{If one replaces the rule (c) with the new rule (c'):
put a bond only if the number $c$ of clusters is kept constant (this was the 
rule chosen in ref.\cite{Gliozzi:2002ub} in the case $Q>1$) then one 
finds $a=p/Q$ and
$b=p$.}. 

This Monte Carlo method with its variants was already  used  to locate the 
marginal value of $Q>2$ in 
three dimensions \cite{Gliozzi:2002ub} and in the study of the backbone 
exponent of critical Q-state Potts models in two dimensions \cite{dbn}.
 
Since this kind of algorithms implies a random sequence of disordering moves 
of type (a) and (b), randomly distributed over the lattice, it led to 
conjecture that they do not suffer of critical slowing down 
\cite{s,Gliozzi:2002ub}.
A subsequent numerical analysis for some integer values of $Q$ in two and 
three dimensions showed that this conjecture is false
\cite{wks}: it reduces critical slowing down, but does not completely eliminate
it, in the sense that its dynamical critical exponent is smaller 
than Swendsen-Wang, but in general does not vanish, at least for $Q>2$ in $2D$
and $Q\ge2$ in $3D$ \cite{wks}. 

Of course, for a practical use of  viable Monte Carlo algorithms it matters 
 not only the intrinsic dynamics of the transition rates, but also
 the efficiency of the numerical implementation. In the present case we 
succeeded in simulating lattice sizes of the order of those used in current 
gauge simulations.
\section{Numerical results}
A simple observable which can be used to locate the threshold 
$p_t=1-e^{-\beta_t}$ where an infinite FK cluster forms is
the connectivity correlator ${\cal{G}}(x,y)=\bra\phi_{x\,y}\ket$, where
$\phi_{x\,y}=1$ only if $x$ and $y$ belong to the same cluster, otherwise
is set to zero. Clearly for $p<p_t$ one observes an asymptotic exponential 
decay  ${\cal{G}}(x,y)\sim e^{-m\,\vert x-y\vert}$ with increasing 
separation $\vert x-y\vert$, where $m$ is the mass of the lowest physical 
state. The correlation length
$\xi=1/m$ is of the order of the mean linear size of the FK clusters. 
 As a consequence, $m$ vanishes at $p_t$, and it is expected to 
obey a critical power law 
\eq
m\simeq a\,(p_t-p)^\nu+b\,(p_t-p)^{\nu'}+\dots
\label{mass}
\en  
where $\nu=\nu(Q)$ is the thermal exponent. Unfortunately it appears that 
this critical 
exponent has not yet been calculated in 3D Q-state Potts models in the range
$0<Q<1$, apart the special $Q\to0$ limit  
(see the second paper of \cite{qzero}). 
\begin{figure}
\begin{flushleft}
{\includegraphics[width=7.5cm]{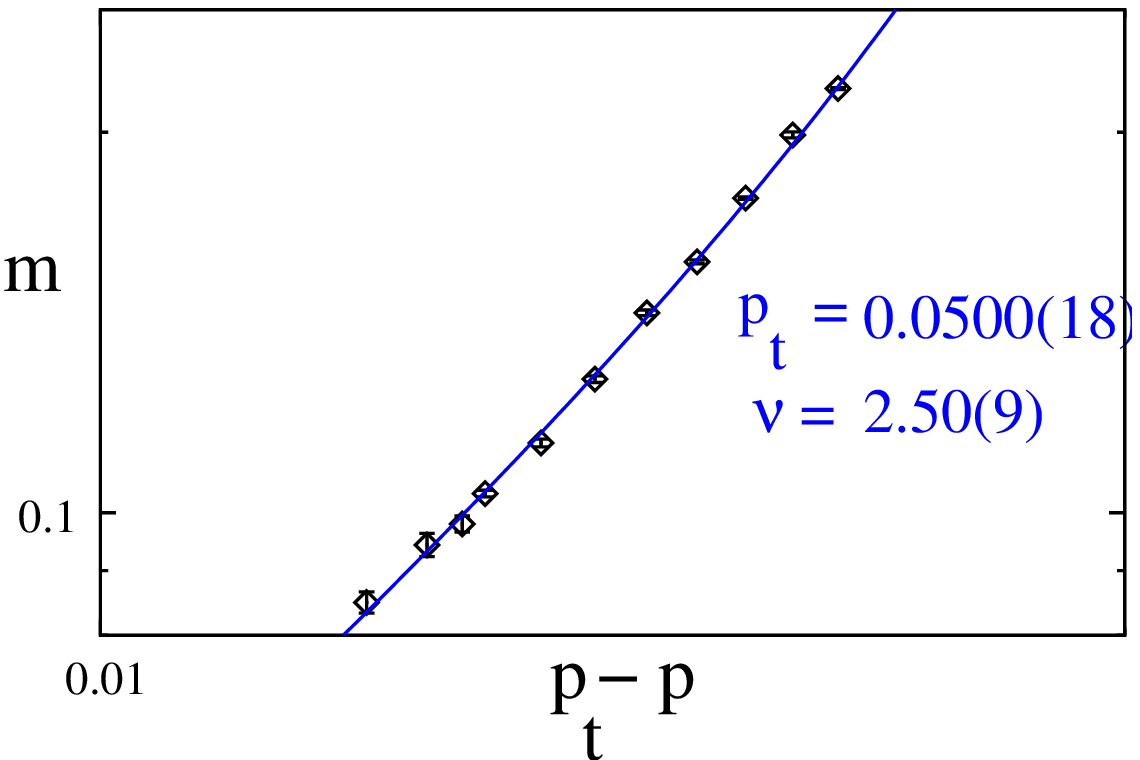}}
\end{flushleft}

\begin{flushright}
\vskip -5.5 cm
{\includegraphics[width=7.5cm]{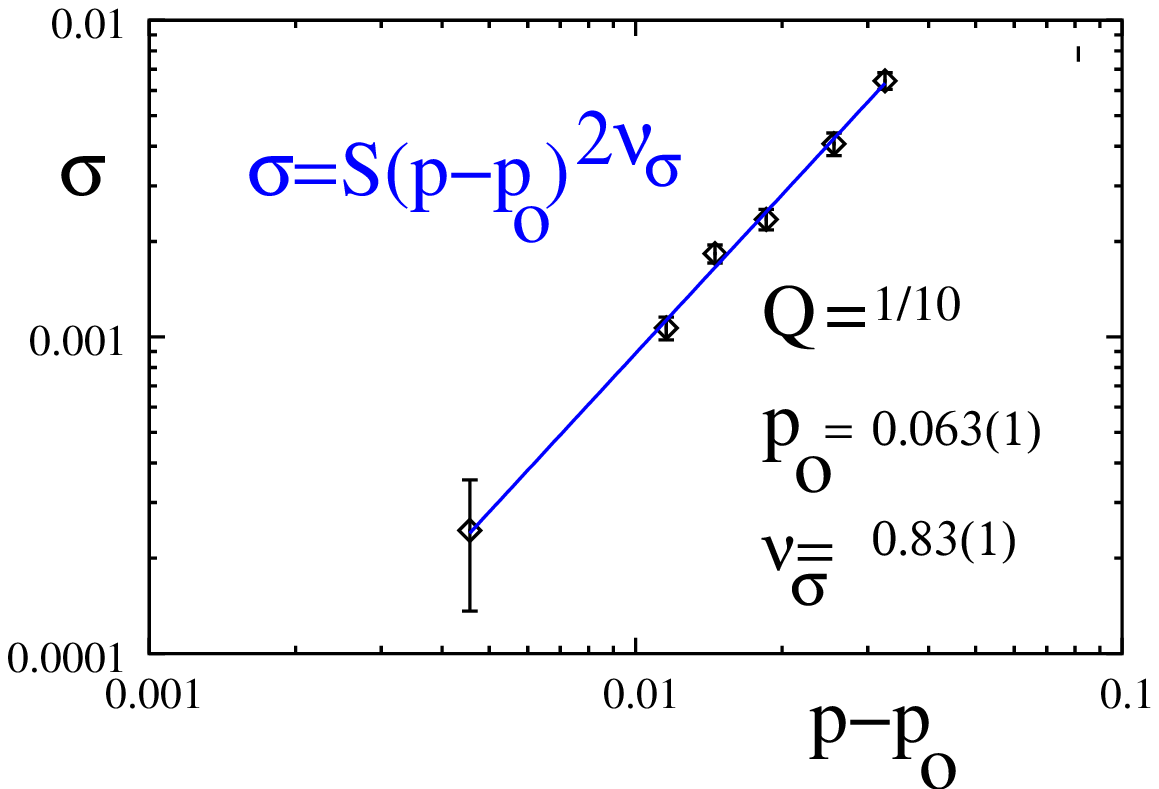}}
\end{flushright}
\caption{The mass of the lowest physical state in the symmetric vacuum (left) 
and the string tension in the confining vacuum(right).  The data refer to 
$Q=1/10$ Potts model in a lattice of size $32^3$.  }
\end{figure}

We performed our simulations on a 
$32^3$ cubic lattice at $Q=\frac1{10}$. We extracted the mass of the lowest 
physical state using the zero momentum projection and fitted the data to 
(\ref{mass}) as shown in Fig.2 (left). As a result the threshold value for 
the formation of a percolating FK cluster and the thermal critical exponent 
turn out to be $p_t=0.0500(18)$ and $\nu=2.50(9)$. 
Note that this value of $\nu$ is much larger than the corresponding value 
at $Q=1$ $\nu_{Q=1}=0.874(2)$. 
This agree with the fact that in two space dimensions the presumed 
exact value of $\nu$ increases as $Q$ decreases. 

On the same lattice at the same value of $Q=\frac1{10}$, but at larger values 
of $p$, where one can easily observe percolation of the sub-clusters made with 
the circuits of the FK clusters, we measured the vacuum expectation value of a 
set of square Wilson loops, in order to evaluate the string tension. They 
perfectly fitted  the expected asymptotic functional form for the 
confining phase 
(including the log term due to the quantum fluctuations of the underlying 
confining string). The extracted string tension as a function of $p$ is 
nicely described by  a power law (see Fig.2 (right): the fitting curve is a 
straight line in the log log scale). However the vanishing point of the 
string tension, $p_o=0.063(1)$, where the deconfining phase starts, does not 
coincide with the threshold $p_t$. In the range $p_t\le p\le p_o$ the 
vacuum of this theory is characterised by a non-vanishing magnetic 
monopole condensate which is not confining, being $\sigma=0$ there.        

Note that, contrarily to what happens in the gauge duals of $Q\ge1$ Potts 
models,  the critical index $\nu_\sigma$ associated to the vanishing of the 
string tension is totally different from the thermal critical index $\nu$. 
It is worth observing that there is no local order parameter that can 
signal when an infinite cluster of circuits forms , being a  
phenomenon of topological nature that can be detected only 
by large Wilson loops. As a consequence  we do not expect that in these 
gauge models the vanishing of the string tension is associated to any sort 
of bulk transition.

\end{document}